\documentclass[fleqn,usenatbib]{mnras}

\usepackage{mathptmx}
\usepackage{txfonts}

\usepackage[T1]{fontenc}
\usepackage{ae,aecompl}
\usepackage{amssymb}    
\usepackage{epsfig}

\newcommand{\coa}{\mbox{$^{13}$CO}}

\newcommand{\kms}{\mbox{km s$^{-1}$}}

\newcommand{\cc}{\mbox{cm$^{-3}$}}
\newcommand{\cmsq}{\mbox{cm$^{-2}$}}
\newcommand{\msun}{\mbox{M$_\odot$}}

\newcommand{\vlsr}{\mbox{$V_{{\rm LSR}}$}}

\title[Short title, max. 45 characters]{MNRAS \LaTeXe\ template -- title goes here}

\title[
Fragmentation of molecular clumps in the Galaxy]{Early science with the Large Millimetre Telescope: 
Fragmentation of molecular clumps in the Galaxy}
\author[M. Heyer, G.W. Wilson et al.]
{M.~Heyer$^1$\thanks{Email:heyer@astro.umass.edu}, 
G.~W.~Wilson$^1$, 
R.~Gutermuth$^{1}$, 
S.~Lizano$^2$, 
A.~Gomez-Ruiz$^3$, \and 
S.~Kurtz$^2$, 
A.~Luna$^3$, 
E.~O.~Serrano~Bernal$^3$, 
and Schloerb, F.P.$^1$
\\
$^{1}$Department of Astronomy, University of Massachusetts, Amherst, MA 01003, USA\\
$^{2}$Instituto de Radioastronom\'ia y Astrof\'isica, Universidad Nacional Aut\'onoma de M\'exico, 58089 Morelia, 
Michoac\'an, M\'exico \\
$^{3}$CONACYT-Institut Nacional de Astropf\'isica, \'Optica y Electr\'onica, Luis E. Erro 1, 72840 Tonanzintla, Puebla, 
M\'exico \\
}
\date{Accepted XXX. Received YYY; in original form ZZZ}

\pubyear{2017}

\begin{document}
\label{firstpage}
\pagerange{\pageref{firstpage}--\pageref{lastpage}}
\maketitle


\begin{abstract}
Sensitive, imaging observations of the $\lambda$1.1~mm dust continuum emission from a 1~deg$^2$ area 
collected with the AzTEC bolometer camera on the Large Millimeter Telescope
are presented.  A catalog of 1545 compact sources is constructed based on a Wiener-optimization filter.  These 
sources are linked to larger clump structures identified in the Bolocam Galactic Plane Survey.  Hydrogen column densities 
are calculated for all sources and mass and mean volume densities are derived for the subset 
of  sources for which kinematic distances can be assigned.  The AzTEC sources are localized, high density 
peaks within the massive clumps of molecular clouds and comprise 5-15\% of the clump mass.  
We examine the role of the gravitational instability in generating these fragments by comparing 
the mass of embedded AzTEC sources to the Jeans' mass of the parent BGPS object.  
For sources with distances less than 6~kpc
the fragment masses are comparable to the 
clump Jeans' mass, despite having isothermal Mach numbers between 1.6 and 7.2.  
AzTEC sources linked to ultra-compact HII regions have mass surface densities greater than 
the critical value implied by the mass-size relationship of infrared dark clouds with 
high mass star formation  while AzTEC sources 
associated with Class II methanol masers 
have mass surface densities greater than  0.7 g cm$^{-2}$ that approaches the proposed threshold required to form massive stars. 
\end{abstract}
\begin{keywords}
ISM:clouds -- ISM:molecules -- ISM: structure -- stars:formation -- submillimetre:ISM
\end{keywords}

\section{Introduction}                                               
The density structure of a molecular cloud is a roadmap for recent and impending 
sites of star formation within its domain.  New stars develop within gravitationally 
unstable protostellar cores in which most of the available gas is subsumed into the 
star or circumstellar disk \citep{Krumholz:2007, Andre:2014}.   The protostellar cores reside within larger 
structures such as clumps and filaments that have mean densities of 10$^{3-5}$ \cc, sizes 
between 0.5 and several pc, and masses ranging from 100 to 10$^4$ \msun\ \citep{Andre:2014,Svoboda:2016}. 
These features are embedded 
within larger molecular clouds that span 10-50~pc in size, have mean volume densities of 10$^{2-3}$ \cc, 
and masses ranging from 10$^4$ to 10$^6$ \msun\ \citep{Heyer:2015}. 
Understanding the processes responsible for cloud fragmentation from clouds to clumps to protostellar 
cores is essential to developing 
improved descriptions of star formation in galaxies.  An important issue to address is 
the efficiency with which 
mass is redistributed into ever more dense gas configurations in each stage of fragmentation. 
This efficiency can regulate the star formation rate in molecular clouds and may also 
impact the distribution of stellar masses of the newborn stars. 

Recent imaging surveys of the dust continuum emission along the Galactic plane illustrate 
the complex structure of the dense molecular interstellar medium and provide a broad 
census of protostellar cores and clumps throughout the Milky Way. 
The Apex Telescope Large Area Survey of the Galaxy (ATLASGAL)
surveyed the 870\micron\ dust emission between longitudes -60 to 60 degrees \citep{Schuller:2009}.  
The Bolocam Galactic Plane Survey (BGPS) imaged the $\lambda$1.1~mm dust emission 
from the Central Molecular Zone (CMZ),
the first quadrant of the 
Galaxy, and targeted star forming regions 
in the Outer Galaxy \citep{Ginsburg:2013}.  
The Herschel Infrared 
Galactic Plane Survey (HI-GAL) imaged the dust emission in five far infrared wavelength bands between 
70 and 500\micron\ \citep{Molinari:2016}.   These programs have 
identified thousands of compact sources that include dense prestellar and protostellar 
cores and massive clumps at angular resolutions 
between 19\arcsec\ and 35\arcsec.  To examine the structure within a dust source that represents 
the next stage of fragmentation, one requires higher angular 
resolutions typically afforded by mm-wave interferometers.  However, owing to the small field-of-view of 
interferometers,
the number of targets one can investigate is rather limited and subject to selection bias.

With the operation of the Large Millimeter Telescope (LMT) in Mexico that is 
equipped with the AzTEC bolometer array at $\lambda$1.1~mm, 
one can collect high sensitivity imaging observations of the dust continuum emission over large areas of the sky 
at 8.5\arcsec\ angular resolution.   Such imaging offers an unbiased census of cold, compact cores that may constitute the 
next level of cloud fragmentation in the interstellar medium (ISM). 
Here, we report on results from imaging of the $\lambda$1.1~mm dust continuum emission over a 1 deg$^2$ field centered at the 
Galactic longitude of 24$^{\circ}$.5 with the LMT.  This line of sight through the 
Galaxy crosses the stellar bar, the near and far sides of both the Sagitarrius and Scutum spiral arms and the 
far side Perseus and 
Norma-Cygnus arms.  Therefore, the field covers a broad range of Galactic radii and interstellar conditions. 
The LMT data are complemented with measurements from the recent Galactic plane surveys of the 
dust and radio continuum emissions, and molecular line emission. 

\section{Data}
The 32~meter Large Millimeter Telescope Alfonso Serrano was used with the AzTEC camera \citep{Wilson:2008} to 
image the $\lambda$1.1~mm dust continuum emission from $\sim$1 deg$^2$ area centered on the Galactic coordinates (24.5, 0).  
The half-power beam width at this wavelength is 8.5\arcsec.
The observations were carried out in Spring 2016 as part of the Early Science-4 cycle of the LMT.  Data were collected 
using On-the-Fly (OTF) mapping in which the telescope is rapidly scanned in azimuth while continuously 
collecting data.  Thirteen individual maps were collected over several nights of observations.  The mean atmospheric 
opacity was 0.23.   The AzTEC calibration procedure is described in \citet{Wilson:2008}.  Accounting for optical 
loading, bolometer responsivities, extinction corrections, and flux conversion factors for each bolometer,
the fractional calibration error over a single night of observing is approximately 10\% and 
decreases in quadrature over multiple nights of data collection.
Pointing measurements on bright, nearby point sources are made approximately every hour throughout the night. 
Positional errors relative to a recent pointing measurement are typically 1\arcsec.

The data were processed with the AzTEC pipeline to remove atmospheric contributions, 
to account for pointing corrections,
to calibrate the data, and to coadd the ensemble of maps \citep{Scott:2008}.  A noise map is constructed from 
jack-knifed noise realizations of each AzTEC map in which 
the time stream is randomly multiplied by $\pm$1.  The effect of the jack-knife step is to 
remove sources while retaining 
noise properties.  Each jack-knifed time-stream is converted into a map and coadded in weighted quadrature  
to produce a final image of rms noise values, $\sigma$.  
The final images of surface brightness and weights (1/$\sigma^2$) 
are constructed in equatorial coordinates (J2000).  To compare with 
other data sets, the images are rotated into Galactic coordinates with the angle -62$^\circ$.6 that is appropriate 
for this segment of the 
Galactic plane.   

\section{Results}
\begin{figure*}
\begin{center}
\epsfxsize=18cm\epsfbox{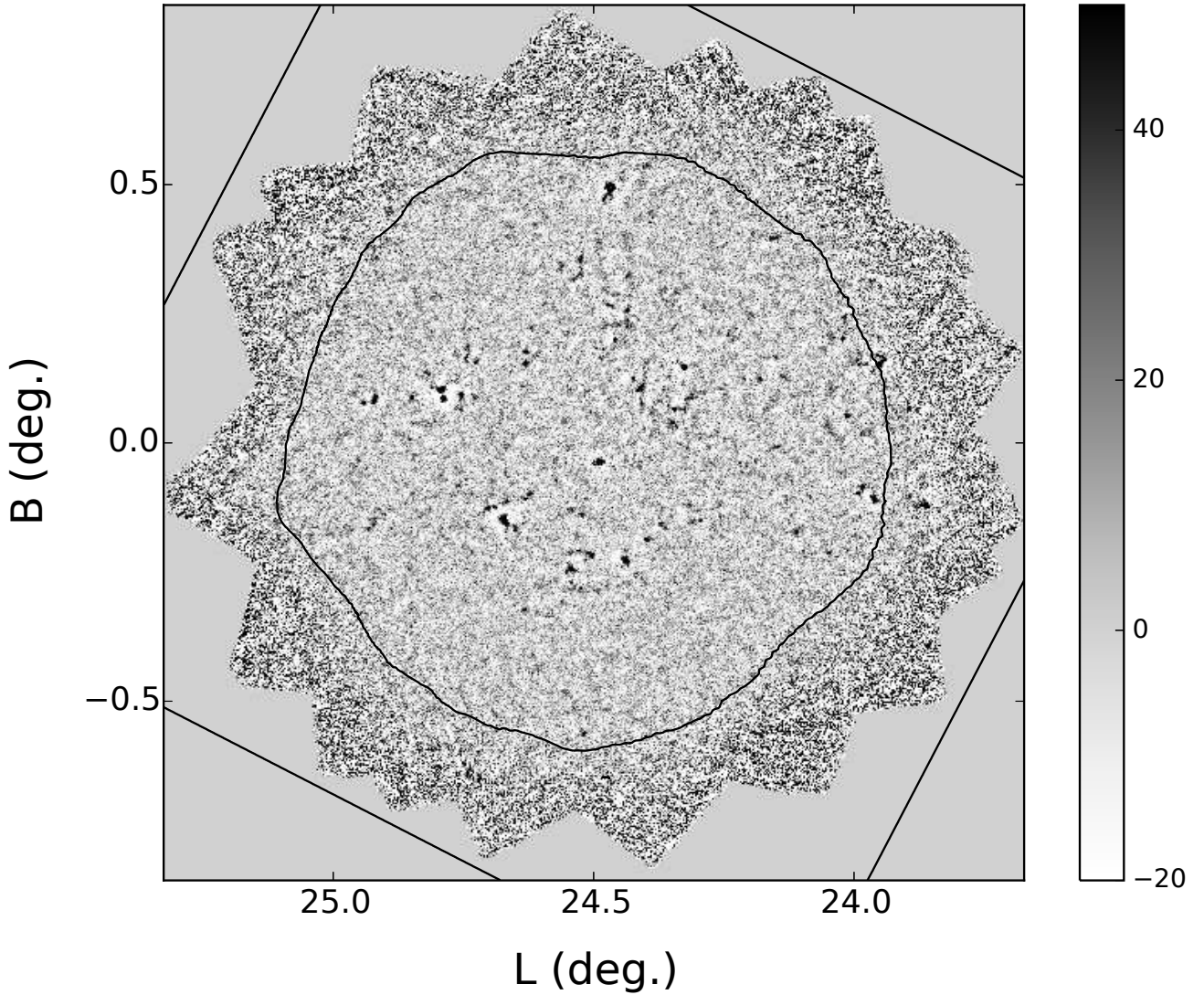}
\caption{
Image of $\lambda$1.1~mm surface brightness over the observed field in units of mJy/beam.
The contour marks the loci at which the rms image  
encloses 70\% 
of the pixels.  
}
\label{fig1}
\end{center}
\end{figure*}

The primary data products from the AzTEC pipeline are an image of $\lambda$1.1~mm surface brightness and a 
corresponding noise image that reflects both instrumental and atmospheric contributions to the noise budget.
Figure~\ref{fig1} shows the image of surface brightness 
rotated to the Galactic coordinate system. 
As with all OTF maps, the outer edges of the field have larger errors owing to 
fewer number of collected samples at these positions, and therefore, less accumulated time,
 relative to the central regions.  
Excluding these outer edges, the median surface brightness 
rms uncertainty is 9 mJy/beam with little variation within the boundary that encloses 70\% of the data 
(solid line in Figure~\ref{fig1}).
This boundary approximates a circle with radius of 0.55 degree centered on l,b=24.5,0.  We restrict 
our analysis to the area within this circle. 

Within the processed image, any smoothly distributed emission component over angular scales greater than $\sim$50\arcsec\ 
is removed along with contributions from the atmosphere. 
This angular-dependent filtering is common to all ground-based millimeter and submillimeter camera systems with the 
filtering scale comparable to approximately half of the angular size of the field-of-view of the camera 
\citep{Ginsburg:2013}.  
The effect of removing the 
extended emission component is that the residual signal is distributed into compact 
structures that are 
sites of higher column density and/or dust temperature with respect to the extended, interstellar cloud. 
These residual features generally correspond to the 
dense clumps and compact cores of molecular clouds.

The surface brightness image in Figure~\ref{fig1} shows several 
distinct clusters of compact emission features distributed throughout the field.  
Several of these clusters are associated with compact HII regions identified in radio 
continuum surveys (see \S3.5).  Members of an 
apparent cluster of AzTEC sources may be physically associated, residing in the same molecular 
cloud complex.  However, additional velocity information for each member is required to more firmly establish this link. 
We note that different clusters and clouds in the field are located at varying 
distances and Galactic radii. 

\begin{figure*} 
\begin{center}
\epsfxsize=18cm\epsfbox{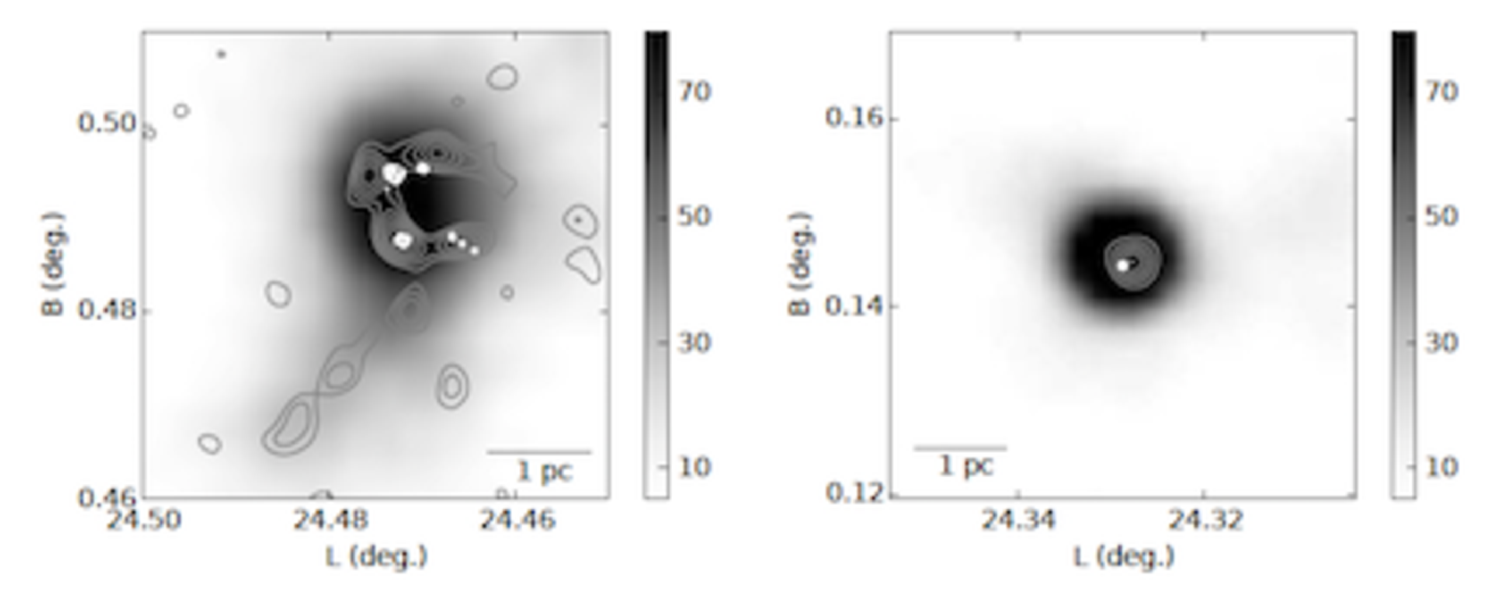} 
\caption{(left) The $\lambda$1.1~mm dust continuum emission observed by the BGPS (halftone image) and AzTEC (grey contours) 
from the environment of a cluster of ultra-compact HII regions.
Units are in MJy/steradian.  
The AzTEC contour levels range from 10-70 MJy/steradian spaced by 10 MJy/steradian.  The white 
contours are 5~GHz radio continumm emission from CORNISH \citep{Purcell:2013}.
(right)  BGPS and AzTEC images of $\lambda$1.1~mm continuum emission from the environment of the Class II 
 methanol maser, G24.329+0.144 \citep{Breen:2015}. The solid white circle marks the location 
of the maser.  The AzTEC contour levels range from 50 to 550 MJy/steradian spaced by 100 MJy/steradian. 
}
\label{fig2}
\end{center}
\end{figure*}

The image of observed surface brightness over the field 
does not effectively convey the measured structure within the image.
The field is effectively 0.95 deg$^2$ in area while the detected signal is distributed into compact objects 
with angular sizes less than 50\arcsec.  Figure~\ref{fig2} shows two subfields 
that illustrate the distribution of 
dust continuum emission detected by AzTEC over smaller angular scales following the application of the Wiener filter.
The left image shows multiple, compact AzTEC sources (grey contours) within the 
domain of the BGPS emission at 34\arcsec\ resolution,
indicating fragmentation of this larger segment of a cloud.  It also demonstrates 
the ability of the improved angular resolution of AzTEC on the LMT to 
more precisely locate the parcel of gas linked to the embedded 
ultra-compact HII (UCHII) regions that are highlighted in white contours.  
The right image shows a singular, bright 
AzTEC source embedded within a marginally resolved BGPS source.   In this case, no fragmentation is evident 
at the 8.5\arcsec\ resolution of the AzTEC data. 
 The peak dust emission is coincident with the 
Class II methanol maser G24.329+0.144 \citep{Breen:2015} that
 marks a very early phase 
of a developing massive star 
\citep{Menten:1991}.  

\subsection{Source Extraction}
The improved sensitivity and angular resolution of the AzTEC data with respect to previous surveys of millimeter dust continuum 
along the Galactic plane motivates the construction of a compact source list.   
To identify compact objects within the data, we apply the source extraction algorithm described 
by \citet{Perera:2013}, which is optimized for unresolved or marginally resolved features within the map.  
In brief, the functional form of the point spread function (PSF) is fit to each pixel in the image, which acts as 
an effective noise-reduction Wiener filter.  
Localized 
peaks in the resultant 
signal-to-noise image correspond to candidate point sources. In practice, the source may be unresolved along one direction but 
marginally extended along the orthogonal axis.  

The source identification algorithm identifies 1545 source-candidates in the observed field with signal to noise greater than 3. 
Source positions in equatorial and Galactic coordinates, peak fluxes, uncertainties, and signal to noise (s2n) 
are derived from the fit of the PSF to the brightness distribution 
 for all identified sources and 
are listed in Table~\ref{table1}. 
The median of the distribution of derived errors for the peak flux values is 7~mJy. 

\begin{table*}
\centering

\caption{Identified AzTEC Sources. \label{table1}
The full table is 
available in the electronic version of the paper.}
\begin{tabular}{rcccccccc}
 \hline
AzTEC ID & Name &  l & b & RA & Dec & S$_{1.1}$ & $\sigma({\rm S}_{1.1})$ & log(P$_J$)\\
     &  & (deg.) & (deg.) & (deg.) & (deg.) & (Jy) & (Jy) &  \\
 \hline
   1 & LMT\_024.302+000.537 &  24.3024 &   0.5367 & 278.4203 &  -7.4269 & 0.031 & 0.010 & -0.62\\ 
   2 & LMT\_024.268+000.515 &  24.2682 &   0.5146 & 278.4242 &  -7.4674 & 0.030 & 0.010 & -0.97\\ 
   3 & LMT\_024.358+000.557 &  24.3581 &   0.5573 & 278.4277 &  -7.3680 & 0.028 & 0.009 & -0.33\\ 
   4 & LMT\_024.379+000.560 &  24.3791 &   0.5601 & 278.4350 &  -7.3480 & 0.029 & 0.009 & -0.38\\ 
   5 & LMT\_024.344+000.539 &  24.3444 &   0.5390 & 278.4378 &  -7.3886 & 0.031 & 0.009 & -1.46\\ 
   6 & LMT\_024.238+000.483 &  24.2376 &   0.4831 & 278.4382 &  -7.5091 & 0.033 & 0.009 & -0.62\\ 
   7 & LMT\_024.388+000.557 &  24.3884 &   0.5568 & 278.4423 &  -7.3413 & 0.029 & 0.009 & -1.28\\ 
   8 & LMT\_024.204+000.459 &  24.2042 &   0.4595 & 278.4437 &  -7.5497 & 0.033 & 0.009 & -0.73\\ 
   9 & LMT\_024.247+000.481 &  24.2473 &   0.4812 & 278.4443 &  -7.5013 & 0.032 & 0.009 & -0.51\\ 
  10 & LMT\_024.313+000.514 &  24.3130 &   0.5140 & 278.4456 &  -7.4280 & 0.028 & 0.009 & -0.86\\ 
 \hline
\end{tabular}
\end{table*}

As part of the source identification process, 20 separate noise realizations of the data 
are searched for sources with the same algorithm to 
provide a measure of the expected number of false detections as a function 
of signal to noise.  
This analysis suggests that 30-50\% of the sources with s2n$<$3.75 within the 
field are 
false detections.  This fraction attenuates with increasing signal to noise 
until limited by Poisson statistical errors.  To quantify the probability of 
a false detection, we fit the expected fraction to an exponential decay curve 
for bins with s2n$<$5.25 that are not limited by Poisson noise, 
\begin{equation}
{\rm P}_{\rm F}(s2n)={\rm P}_{\rm F}(3){\rm exp}(-\alpha({\rm s2n}-3.0)). 
\end{equation}
The best fit values are P$_F$(3)=0.47 and $\alpha$=0.61.
Equation 1 describes a statistical expection of the fraction of sources in each bin 
that are false.  It does not identify or select which sources are false. 
One could simply exclude low signal to noise sources from further analysis but 
this selection criterion would also remove a significant number of valid sources. 

To refine our measure of spurious sources, we construct a secondary 
probability function based on the 
angular distribution of $\lambda$1.1~mm surface brightness independently measured by 
the BGPS that is more effective at recovering extended dust emission from clumps and clouds.   
Compact dust sources associated with potential or active sites of star formation 
are expected to reside within the boundaries 
of molecular clouds and clumps while false sources and contaminants such as background galaxies 
should be randomly distributed 
within the field. 
The probability that a randomly selected position in the field is coincident with a BGPS pixel (7.2\arcsec\ in size)
with surface brightness greater than $I$ is equal to the area subtended 
by these pixels normalized by the total area or equivalently, 
\begin{equation}
{\rm P}(I_{BGPS}>I)=\psi(I_{\rm BGPS}>I)/\psi_{T}
\end{equation}
where $\psi(I_{\rm BGPS})$ is the number of BGPS pixels with 
surface brightness, $I_{BGPS}$, 
and $\psi_T$ is the total number of BGPS pixels in the selected area. 
To evaluate equation 2, only BGPS pixels within the 0.55 degree radius are 
included. 

\begin{figure*} 
\begin{center}
\epsfxsize=18cm\epsfbox{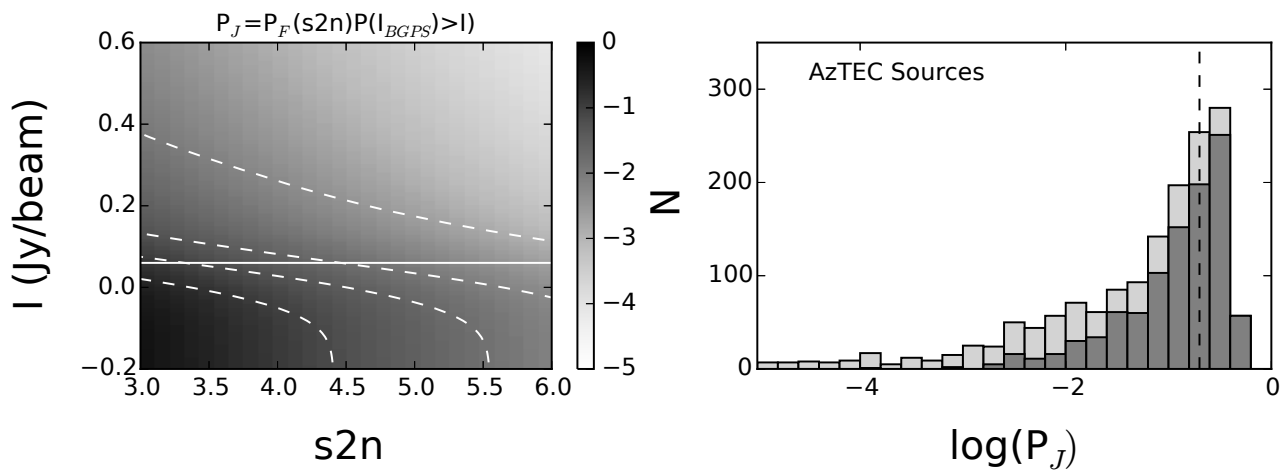} 
\caption{ (left) The logarithm of the joint probability, ${\rm P}_{\rm J}$, of finding a false source with a 
given signal to noise to be 
coincident with a BGPS pixel with surface brightness greater than $I$ within the observed field. 
The dotted contours show ${\rm P}_{\rm J}$ values of 0.01,  0.05, 0.1, and 0.2 from top to bottom.
The solid line is the median rms noise level of the BGPS for this field \citep{Ginsburg:2013}.
(right) Distribution of log(${\rm P}_{\rm J}$) for all identified AzTEC sources (light-grey) and for sources 
with s2n $<$ 3.5 (dark-grey).  
The vertical dotted line marks the joint probability threshold adopted by this study to select
reliable sources, ${\rm P}_{\rm J} <$ 0.2. 
}
\label{fig3}
\end{center}
\end{figure*}

From these two independent probability functions (equations 1 and 2) 
 that a source is falsely identified as a compact, Galactic dust source,
we calculate the 
joint probability distribution, 
\begin{equation}
{\rm P}_{\rm J}={\rm P}_{\rm F}(s2n){\rm P}(I_{BGPS}>I).
\end{equation}
Figure~\ref{fig3} (left) 
shows the variation of log(${\rm P}_{\rm J}$) in the s2n-I plane.  The drawn dotted contours 
represent ${\rm P}_{\rm J}$ values 0.01, 0.05, 0.1, 0.2 from top to bottom.  The horizontal 
solid line denotes the 1$\sigma$ rms of 60 mJy/beam for the BGPS surface brightness in this field 
\citep{Ginsburg:2013}. 

For each 
AzTEC source, 
we evaluate ${\rm P}_{\rm J}$ that effectively measures the probability that 
the source is false given its signal to noise ratio and location in the map with respect to the 
BGPS surface 
brightness distribution.  
These ${\rm P}_{\rm J}$ values are listed in 
Table~\ref{table1} for each source 
and can be used to select sources with increasing reliability (lower ${\rm P}_{\rm J}$). 
Figure~\ref{fig3} (right) shows the 
distribution of log(${\rm P}_{\rm J}$) values for all AzTEC sources 
(light grey histogram) and for sources with s2n$<$3.5 (dark grey histogram) 
that comprise 
most of the expected false detections.  

Decreasing values of ${\rm P}_{\rm J}$ reflect higher confidence that a given source is not false.
For this study, we choose a limiting ${\rm P}_{\rm J}$ value of 0.2.  The number of sources 
excluded by this ${\rm P}_{\rm J}$ value approximates the number of expected sources to be false 
for a given s2n value
but also 
considers the added information provided by BGPS imaging.   There are 1095 sources with $P_J<$0.2.
This selection does not eliminate all low 
signal to noise sources but only those that fall into regions of low surface 
brightness as measured by the BGPS.  Conversely, high signal to noise sources 
(s2n$>$4.5) 
that are coincident with low BGPS surface brightness signal are considered valid and likely reflect the increased 
source sensitivity of the LMT data relative to the BGPS. 
This selection of sources can lead to a bias in which sources are spatially 
clustered but such a state is physically expected given the limited solid 
angle of molecular clouds and dense clumps in the ISM. 

\subsection{Linking AzTEC Sources to BGPS Sources}
It is useful to compare sources identified by these AzTEC measurements with those of previous dust continuum imaging 
surveys of the Galactic plane measured with lower angular resolution.  
The BGPS is especially convenient as its detector bandpass is identical 
to that of AzTEC so no assumptions of the dust emissivity are required to make quantitative comparisons.
Moreover, there 
is a large amount of auxiliary information related to the BGPS sources including spectroscopy 
of high density gas tracers 
\citep{Dunham:2011, Schlingman:2011, Shirley:2013, Svoboda:2016} and kinematic distance assignments
\citep{Ellsworth:2015a, Svoboda:2016}.
This added information can be applied to the set of AzTEC sources that are linked to a given BGPS source.  

To make this link,
we use the image mask product provided by the BGPS Version 2.1 
that marks the observed pixels for each BGPS source \citep{Ginsburg:2013}.  
An AzTEC source 
is directly linked if its position falls within the solid angle of a BGPS source defined by the image mask.  
Table~\ref{table2} lists the BGPS version 2 catalog number for each 
directly linked AzTEC source.
We acknowledge that this method is not infallible.  A fraction of the linked AzTEC sources could be foreground or background to the BGPS source.  
Ideally, one would require a spectroscopic velocity measurement of a dense gas tracer 
for each AZTEC source to directly 
relate it to the velocity-tagged BGPS 
source. 

\begin{table*}
\centering

\caption{AzTEC Sources Linked to BGPS Objects.\label{table2}
Values of -999.0 indicate no velocity or distance measure is available.  
Column 7: the kinematic distance ambiguity resolution (kdar) flag -- N (near), F (far), U (unknown).
Column 8: source of dense gas velocity and distance measure--
0 (unknown) 1 \citep{Ellsworth:2015a}, 2 \citep{Svoboda:2016}, 3 proximity to BGPS 
source with known distance, 4 \citep{Roman-Duval:2010}, 5 (other).
The full table is 
available in the electronic version of the paper}
\begin{tabular}{rccccccc}
 \hline
AzTEC ID & BGPS CNUM & V$_{\rm LSR}$ & D &  -$\sigma(D)$ &  +$\sigma(D)$ &  kdar & Ref\\
         &           &  (\kms)       & (kpc) & (kpc) & (kpc) & \\
 \hline
  22 & 4058  &   93.6 &   10.1 &    0.2 &    0.2 & F & 2 \\
  26 & 4058  &   93.6 &   10.1 &    0.2 &    0.2 & F & 2 \\
  27 & 4067  &   96.3 &    5.1 &    0.3 &    0.3 & N & 1 \\
  29 & 4058  &   93.6 &   10.1 &    0.2 &    0.2 & F & 2 \\
  31 & 4040  & -999.0 & -999.0 & -999.0 & -999.0 & U & 0 \\
  33 & 4040  & -999.0 & -999.0 & -999.0 & -999.0 & U & 0 \\
  34 & 4067  &   96.3 &    5.1 &    0.3 &    0.3 & N & 1 \\
  39 & 4151  &  102.2 &    5.2 &    0.2 &    0.2 & N & 2 \\
  42 & 4116  &  102.0 &    5.3 &    0.2 &    0.2 & N & 2 \\
  44 & 4151  &  102.2 &    5.2 &    0.2 &    0.2 & N & 2 \\
 \hline
\end{tabular}
\end{table*}

A comparison of the distribution of AzTEC sources with the BGPS objects shows several cases.  
First, there are 619 AzTEC sources with $P_J<$0.2 that are linked to 
BGPS objects.  
Of the 255 BGPS catalogued objects within 0.55 degrees of the map center, 
there are 65 BGPS targets with a single AzTEC source and 124 BGPS targets with two or more AzTEC sources.
This multiplicity within BGPS defined clumps is discussed further in \S4. 
Second, 
there are 
66 BGPS clumps 
with no AzTEC counterparts with ${\rm P}_{\rm J}<0.2$.  
Given the improved sensitivity of the AzTEC data, the absence of any 
AzTEC sources would seem peculiar.
We have visually inspected the BGPS images of these sources.  In a few cases, the 
BGPS sources may be false positive detections.
More typically, one finds low surface brightness emission  
with no significant substructure on scales less 
than the angular extent 
of the AzTEC footprint of $\sim$50-100\arcsec.  In this case,
the AzTEC processing to remove atmospheric contributions also subtracts any smooth, extended, diffuse component -- leaving little 
or no residual signal to be identified by the AzTEC source extraction algorithm. 
Third, there are 476 
AzTEC sources with ${\rm P}_{\rm J}<0.2$ that have no BGPS counterpart.  This can be attributed to 
the improved point source sensitivity and angular resolution of the AzTEC data.

\subsection{Distribution in the Galaxy}
With the linking of a subset of AzTEC sources to a given BGPS source, 
we poll the BGPS auxiliary
data compiled 
by \citet{Ellsworth:2015a} and \citet{Svoboda:2016} to assign velocities derived from a 
dense gas tracer (HCO$^+$ J=3-2 and NH$_3$ (1,1) inversion transition) 
and, if available,  distances.  
\citet{Ellsworth:2015a} compute a Bayesian distance probability function to resolve 
the near-far side 
distance ambiguity using a set of priors for a large number of BGPS sources.  
\citet{Svoboda:2016} extended the method using NH$_3$ observations.  If available, we adopt these 
measures of distance to a linked AzTEC source.  BGPS sources are frequently clustered spatially and 
kinematically.  If a nearby 
BGPS object with an assigned 
distance is within 102\arcsec\ (3 BGPS half-power beam widths) of an AzTEC-linked BGPS object and its dense gas velocity is within 
3~\kms\ of the linked BGPS object, then this distance is applied to all of the embedded AzTEC sources. 
Velocities and distances from these auxiliary data sets are 
listed in Table~\ref{table2}. 

Not all BGPS objects linked to one or more AzTEC sources 
have an established dense gas velocity or distance. 
In these cases, the \coa\ J=1-0 spectrum from the 
Boston University-FCRAO 
Galactic Ring Survey (GRS) \citep{Jackson:2006} at the AzTEC source position is examined.
Such spectra frequently exhibit multiple 
velocity components as these lines of sight traverse several spiral arm features. 
A secondary velocity is assigned to the \vlsr\ value at which the antenna temperature of the 
coincident \coa\ J=1-0 spectrum is maximum.   This peak temperature likely reflects enhanced column density or 
optically thick emission at this velocity within the 48\arcsec\ beam.   In addition, we calculate 
a contrast value, $C_v$, defined as the ratio of the antenna temperatures at the 
velocity of the brightest \coa\ component to that of the next brightest component that is 
displaced by at least 5 \kms.  $C_v$ 
measures the degree to which the selected velocity component is distinguished from 
other components in the spectrum.  
The larger the contrast, the more likely this 
velocity is properly assigned to the AzTEC source.  
This velocity assignment from a \coa\ spectrum is not as conclusive as a spectrum from a 
dense gas tracer but is available for 
all 1545 AzTEC sources.  
For the 554 AzTEC sources with both a dense gas and \coa\ velocity, 493 (89\%) have 
velocity differences, $|$\vlsr-V($^{13}$CO)$|$ less than 5~\kms.   Restricting the sample to sources with 
$C_v>$1.5 (452 sources), this fraction increases to 95\%. 
The \coa-derived velocities and contrast values are compiled in Table~\ref{table3}.  

\begin{table*}
\centering

\caption{AzTEC \coa\ Velocities and Dust Column Densities.\label{table3}
The full table is 
available in the electronic version of the paper}
\begin{tabular}{rccccc}
 \hline
AzTEC ID & V($^{13}$CO) & $C_v$ & N$_{\rm H}$ & -$\sigma$(N$_{\rm H}$) & +$\sigma$(N$_{\rm H}$) \\
     & (\kms) &     & (10$^{22}$ \cmsq) & (10$^{22}$ \cmsq) & (10$^{22}$ \cmsq)\\
 \hline
   1  &  100.4 &  1.2  & 2.05 & 0.99 & 0.82\\ 
   2  &   98.1 &  1.2  & 2.00 & 0.97 & 0.79\\ 
   3  &  102.1 &  2.5  & 1.85 & 0.93 & 0.78\\ 
   4  &  104.1 & 14.8  & 1.93 & 0.92 & 0.76\\ 
   5  &  102.2 &  1.8  & 2.08 & 0.99 & 0.83\\ 
   6  &   98.3 &  2.0  & 2.11 & 1.01 & 0.83\\ 
   7  &  104.0 &  6.3  & 1.98 & 0.98 & 0.76\\ 
   8  &   50.5 &  1.1  & 2.24 & 1.07 & 0.90\\ 
   9  &   98.1 &  2.2  & 2.13 & 1.01 & 0.82\\ 
  10  &  100.4 &  1.9  & 1.88 & 0.95 & 0.79\\ 
 \hline
\end{tabular}
\end{table*}

If there is no BGPS-based distance information from a dense gas tracer, as in the case where the 
kinematic distance ambiguity could not be resolved, we examine the catalog of molecular clouds 
constructed from the GRS \citep{Roman-Duval:2010}.  If the AzTEC 
source is within the projected area of the cloud and its dense gas velocity 
is within 5~\kms\ of the cloud's \coa-defined velocity, 
then the distance to the AzTEC source is assigned to the cloud's distance. 
The distance, ${\rm D}$, asymmetric distance uncertainties, -$\sigma(D)$, +$\sigma(D)$,
resolution 
of the near-far side distance ambiguity, (near, far, uncertain), and distance reference 
for each linked AzTEC source are listed in Table~\ref{table2}.

\begin{figure*}
\begin{center}
\epsfxsize=18cm\epsfbox{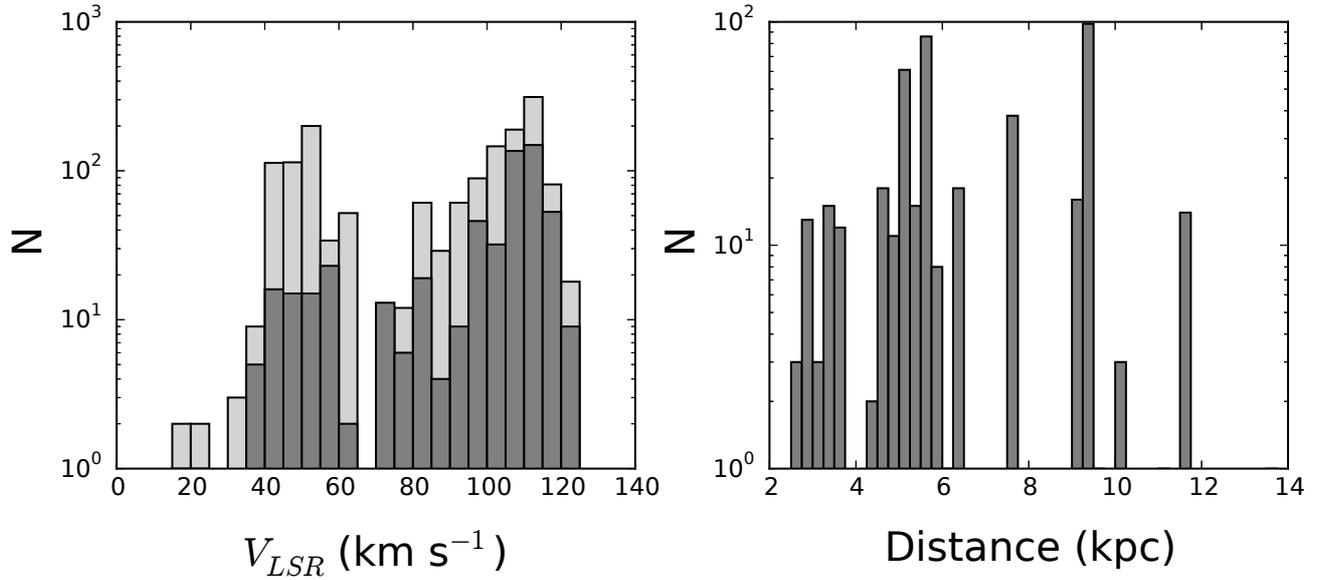}
\caption{(left) Distribution of \vlsr\ values derived from high density gas tracer (dark histogram) and \coa\ J=1-0 
emission (light histogram).
(right) The distribution of distances to the subset of AzTEC sources linked to BGPS objects with defined distances 
from \citet{Ellsworth:2015a} and \citet{Svoboda:2016} or linked to \coa\ clouds catalogued by 
\citet{Roman-Duval:2010}.
}
\label{fig4}
\end{center}
\end{figure*}

The distributions of \vlsr\ values derived from the dense 
gas spectra (dark histogram) and \coa\ spectra (light grey histogram) are shown in Figure~\ref{fig4} (left).
The distributions are similar with peaks at 50~\kms, 80~\kms\ and 110~\kms\ that correspond to the 
expected velocities of the 
far Sagittarius arm, the near Scutum arm and 
the far Scutum arm 
respectively \citep{Steiman:2010, Koda:2016}.  
Figure~\ref{fig4} also shows the 
distribution of distances to AzTEC sources directly or indirectly linked to BGPS objects or linked to 
GRS \coa\ clouds.
Most of these sources are located on the near side of the terminal velocity distance of 7.6~kpc at l=24.5 
assuming the rotation curve of \citet{Reid:2009}.
However there are several clusters of sources at distances 9.25, 10, and 11.5~kpc located on the far side of the 
Galaxy. 


\subsection{Column density, mass, and mean volume density}

The derived properties of mass, mean volume density, and column density are useful measures to 
classify millimeter sources of dust continuum emission into clouds, clumps, and cores 
\citep{Dunham:2011}.  
There are obvious dependencies on distance for mass (D$^2$) and volume density (D$^{-1}$) but these trends 
are also a product of the effective spatial filtering as part of the removal of the atmospheric contribution to 
the signal that depends on the angular scale of the cloud or clump \citep{Battisti:2014}.  For the BGPS sample, the effect is to generally identify more nearby objects as 
cores, sources at intermediate 
distances as clumps, and more distant objects as clouds \citep{Dunham:2011, Ellsworth:2015b}.

To gain a better understanding of the nature of the identified AzTEC compact sources, 
we calculate the hydrogen column 
density for all detected objects and molecular hydrogen mass and mean volume density for the subset of objects with a defined 
distance.  The dust emission at $\lambda$1.1~mm is assumed to be optically thin, which allows a simple 
relationship between hydrogen column density, N$_{\rm H}$, and measured surface brightness, S$_{\rm peak}/\Omega_{\rm A}$,
\begin{equation}
{\rm N}_{\rm H} =\left(\frac{{\rm S}_{\rm peak}}{\Omega_{\rm A}}\right) \frac{{\rm R}}{\mu m_{\rm H}\kappa_{1.1mm} {\rm B}_\nu(T_{\rm D}) }
\end{equation}
where $\Omega_A$ is the solid angle of the AzTEC beam,  R is the gas to dust mass ratio, 
$\mu$=2.4 is the mean atomic weight, and ${\rm m}_{\rm H}$ is the mass of the hydrogen atom.
$\kappa_{1.1mm}$ is dust opacity at $\lambda$1.1~mm and B$_\nu$(T$_{\rm D}$) is the Planck function evaluated 
at $\lambda$1.1~mm and a single temperature, T$_{\rm D}$ for the dust grains.  
We adopt the conventional values for the constants: R=100, $\kappa_{1.1mm}$=1.14 cm$^2$ g$^{-1}$
\citep{Ossenkopf:1994}.
Without an infrared-mm spectral energy distribution, a dust temperature for each source can 
not be derived. Rather, for each source, 
a 
gaussian distribution of dust temperatures centered on 16~K with a dispersion of 4~K is assumed 
that is based on recent 
Herschel estimates of dust temperatures for a set of infrared dark clouds (IRDCs) \citep{Traficante:2015} 
and the median gas kinetic temperature 
\citep{Svoboda:2016}.  From this distribution, a Monte-Carlo 
calculation is executed that randomly samples the dust temperature distribution as input to equation~4 while also 
considering photometric errors.  For 1024 realizations, the result is a distribution of N$_{\rm H}$ values.  The mean value of this 
distribution is assigned to the column density of the source.  Both lower and upper 1-$\sigma$ errors are derived from the shape of the N$_{\rm H}$ distribution.
These errors are primarily due to the sampling of the temperature distribution.  
The column density and uncertainties are listed in  
Table~\ref{table3} for all AzTEC sources.  

This column density represents an average value over the solid angle of the telescope beam.  As these objects are 
identified as point sources or are marginally resolved, the solid angle of one or more sources within the beam 
is smaller than $\Omega_A$ such that actual source column densities would be larger by the factor equal to the 
reciprocal of the beam filling factor of 
sources. Also, 
our assumed dust temperature distribution may not apply to all objects -- especially those with embedded, 
compact HII regions excited 
by massive stars that can locally heat the dust to temperatures in excess of 16~K \citep{Merello:2015}.
The adoption of a higher dust temperature leads to smaller column densities. 

We acknowledge that free-free radiation from an associated compact HII region may contribute to the observed 
$\lambda$1.1~mm emission. 
\citet{Schloerb:1987} found contributions of free-free emission to the $\lambda$1.3~mm band to be less than 10\% in Orion, 25\% in W49A, 
and dominant in W51~IRS2.  More recently, 
\citet{Zhang:2014} estimate that 25-55\% of the emission at $\lambda$1.3~mm is free-free radiation from the hypercompact 
HII region  G35.58-0.03. 
As discussed in \S3.5, only 14 of the AzTEC sources are associated with a compact HII region so 
contamination of the dust emission by free-free emission is likely to be limited to a small number of sources.
For 11 of these sources, we extrapolate the available 5~GHz flux \citep{Purcell:2013} to 273 GHz assuming a spectral 
index of -0.1.  For those sources whose measured angular extent at 5 GHz is greater than 8.5\arcsec, 
we also apply a correction to account for the flux that would be collected within the AzTEC solid angle.
The estimated fractional contribution of free-free emission is the ratio of this extrapolated flux to that 
measured by AzTEC.  Four of the 11 sources have fractional contributions greater than 50\% and seven 
of the sources have contributions in excess of 20\%.  In these cases, the dust derived column densities,
uncorrected for free-free contamination, are upper limits.

\begin{figure*}
\begin{center}
\epsfxsize=15cm\epsfbox{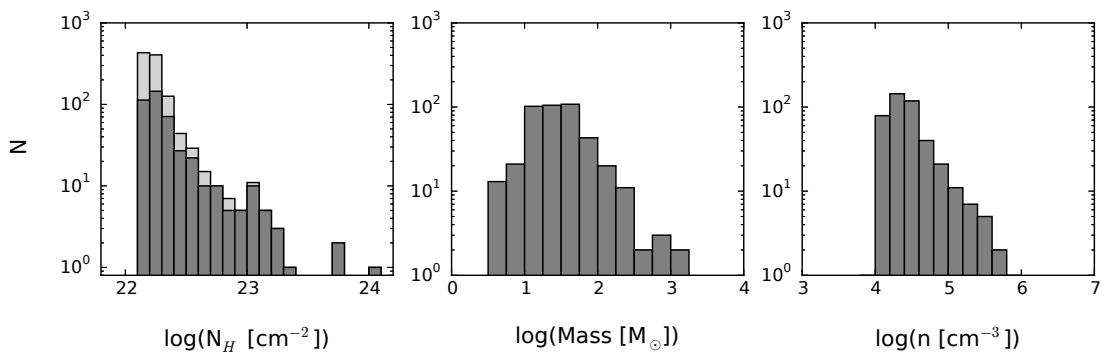}
\caption{ (left) Distribution of the logarithm of hydrogen column density for ${\rm P}_{\rm J}< 0.2$ sources in the field 
(light grey) and those 
with defined distances (dark grey). 
(middle) Distribution of the logarithm of
gas mass within AzTEC sources.  (right) Distribution of the logarithm of mean volume gas density of AzTEC sources. 
}
\label{fig5}
\end{center}
\end{figure*}

The distribution of hydrogen column density for all AzTEC defined sources is shown 
in Figure~\ref{fig5}.    The sharp cutoff for column densities less than 1.3$\times$10$^{22}$ cm$^{-2}$ 
arises from the $>$3$\sigma$ requirement
for source inclusion, the adopted constant for source solid angle, $\Omega_A$,  and 
the uniform sensitivity of the map from which sources are identified.
The  median value of this distribution is 1.7$\times$10$^{22}$ cm$^{-2}$, which is a factor of 4
larger than the weighted BGPS column density integrated over the source area for starless and protostellar 
BGPS sources
derived by \citet{Svoboda:2016} and 2 times larger than their mean-weighted peak column density.
The higher angular resolution of AzTEC 
and the compact definition of the extracted sources lead to the detection of objects with systematically higher 
column densities than typical BGPS sources.  

The mass of material within the AzTEC beam including contributions from helium 
is
\begin{equation}
M=\mu {\rm m}_{\rm H} {\rm D}^2{\rm N}_{\rm H}\Omega_{\rm A} 
\end{equation}
where D is the distance to the source.
The distribution of derived source masses is shown in Figure~\ref{fig5} (middle). 
The masses range from 3.4 to 1625~\msun\ and correspond to  
low mass dense cores that may form a single low-mass star to massive clumps from which several massive stars and 
small stellar clusters could develop.  
Assuming a dust temperature of 16~K, our sensitivity limits can recover masses of compact sources
greater than 10~\msun~(D/4~kpc)$^2$ with signal to noise of 4.  

The mean volume density of each AzTEC source with a defined distance is estimated from the column density and projected 
physical size, which is assumed to be equal to the line of sight depth, 
\begin{equation}
{\rm n}=\frac{{\rm N}_{\rm H}}{\theta {\rm D}}
\end{equation}
where $\theta$ is the AzTEC half-power beam width in radians.  The densities are typical of resolved, dense cores found in nearby star 
forming regions.  Moreover, these are generally higher than mean volume densities derived for 
BGPS sources \citep{Dunham:2011, Battisti:2014, Svoboda:2016}.  

Masses and mean volume densities and 
corresponding 1$\sigma$ uncertainties 
are compiled in Table~\ref{table4}. 
In summary, AzTEC sources have lower masses, higher column and volume densities than BGPS sources and correspond to
localized, high density peaks embedded within a lower density substrate of gas.

\begin{table*}
\centering

\caption{AzTEC Masses and Mean Volume Densities.\label{table4}
The full table is 
available in the electronic version of the paper}
\begin{tabular}{rcccccc}
 \hline
AzTEC ID & M & -$\sigma({\rm M})$ & +$\sigma({\rm M})$ & n & -$\sigma({\rm n})$ & +$\sigma({\rm n})$ \\
         & (M$_\odot$) & (M$_\odot$) & (M$_\odot$) &  (\cc) & (\cc) & (\cc) \\
 \hline
   22  &  134.2 &   54.2 &   44.6  & 31500.0 & 12700.0 & 10400.0 \\ 
   26  &  107.3 &   44.6 &   34.0  & 25200.0 & 10400.0 & 7940.0 \\ 
   27  &   13.8 &    6.9 &    6.3  & 25600.0 & 12600.0 & 11300.0 \\ 
   29  &   67.9 &   31.3 &   25.3  & 16000.0 & 7340.0 & 5930.0 \\ 
   34  &   13.6 &    6.6 &    5.8  & 25200.0 & 12000.0 & 10500.0 \\ 
   39  &   18.9 &    8.7 &    7.3  & 33200.0 & 15000.0 & 12500.0 \\ 
   42  &   23.6 &   10.4 &    8.8  & 39200.0 & 17100.0 & 14400.0 \\ 
   44  &   24.2 &   10.1 &    9.5  & 42600.0 & 17500.0 & 16400.0 \\ 
   46  &   17.2 &    8.2 &    7.2  & 28500.0 & 13400.0 & 11800.0 \\ 
   48  &   14.0 &    6.8 &    5.4  & 23300.0 & 11100.0 & 8810.0 \\ 
 \hline
\end{tabular}
\end{table*}

\subsection{Linking AzTEC sources to compact HII regions and Methanol Masers}
We seek to connect the 
compact, high density regions detected by AzTEC with signatures of massive star formation.  
Such signatures include UCHII regions and Class II 
methanol masers that trace an even earlier stage of forming a massive star. 

For the compact HII regions, the source catalogs from the 
Coordinated Radio and Infrared Survey for High-mass star formation
(CORNISH)
at a frequency of 5 GHz \citep{Purcell:2013} and the HI, OH, Recombination Line Survey of the Milky Way (THOR) at a frequency of 1.4~GHz \citep{Bihr:2016} 
are examined for proximity to a given AzTEC source in the 
field. To isolate sources of free-free emission in the THOR wavelength band of 1.4 GHz, we only consider sources with spectral indices $>$-0.1 derived from 
the sub-bands of the THOR data.
A link between an AzTEC source and a catalogued HII region is established if the distance between the radio continuum 
centroid coordinates 
and the AzTEC peak coordinates is less than 10\arcsec. 
The 14 AzTEC sources linked to HII regions from these catalogs are compiled in Table~\ref{table5}.
\begin{table*}
\centering

\caption{AzTEC Sources Linked HII regions. \label{table5}}
\begin{tabular}{rlcr}
 \hline
AzTEC ID & CORNISH Name  & THOR Name & THOR Spectral Index \\
 \hline
  68 & G024.4698+00.4954 & null & -999.00\\
  77 & G024.4736+00.4950 & null & -999.00\\
  79 & G024.4721+00.4877 & G24.471+0.488 &    0.15\\
 117 & null & G24.677+0.549 &    1.19\\
 150 & G023.9564+00.1493 & G23.956+0.150 &    0.61\\
 263 & null & G24.200+0.192 &    0.13\\
 391 & G024.1839+00.1199 & null & -999.00\\
 573 & null & G23.988-0.089 &   -0.02\\
 923 & G024.4921-00.0386 & G24.493-0.038 &    0.24\\
 965 & G024.7984+00.0967 & G24.798+0.096 &    0.11\\
 988 & G024.7889+00.0824 & null & -999.00\\
1038 & G024.8497+00.0881 & null & -999.00\\
1103 & G024.9237+00.0777 & null & -999.00\\
1219 & G024.5065-00.2224 & G24.507-0.223 &    0.20\\
 \hline
 \hline
\end{tabular}
\end{table*}

A Class II methanol maser from the catalog compiled by \citet{Breen:2015} 
is linked to the AzTEC source 
if its position is within 10\arcsec\ of the AzTEC source position.  
Table~\ref{table6} lists the 14 AzTEC sources that are positionally linked to the methanol masers.
Figure~\ref{fig2} shows examples of AzTEC sources associated with these signatures of massive star formation. 
\begin{table*}
\centering

\caption{AzTEC Sources Linked Class II Methanol Masers. \label{table6}}
\begin{tabular}{rllcr}
 \hline
AzTEC ID & Maser Name  & l(Maser)  & b(Maser) & V(Maser) \\
     &       & (deg.) & (deg.) & (\kms) \\
 \hline
 380 & G24.541+0.312 & 24.5411 &  0.3121 & 106.5\\
 470 & G24.329+0.144 & 24.3287 &  0.1444 & 110.3\\
 494 & G24.461+0.198 & 24.4612 &  0.1980 &  125.5\\
 573 & G23.986-0.089 & 23.9861 & -0.0890  & 65.1\\
 585 & G24.148-0.009 & 24.1479 & -0.0091  & 17.4\\
 599 & G23.966-0.109 & 23.9661 & -0.1092  & 70.9\\
 611 & G23.996-0.100 & 23.9964 & -0.0995  & 68.2\\
 923 & G24.493-0.039 & 24.4934 & -0.0389  & 115.2\\
 988 & G24.790+0.083a & 24.7897 &  0.0832 & 113.3\\
1038 & G24.850+0.087 & 24.8502 &  0.0874 & 110.0\\
1087 & G24.920+0.088 & 24.9196 &  0.0882  & 53.3\\
1121 & G24.943+0.074 & 24.9428 &  0.0741  & 46.6\\
1237 & G24.676-0.150 & 24.6755 & -0.1499 & 116.1\\
1409 & G24.634-0.324 & 24.6340 & -0.3237 &  35.6\\
 \hline
\end{tabular}
\end{table*}

\section{Fragmentation of Massive Clumps}
The BGPS source catalog provides a 
Galactic census of massive clumps that are responsible for clustered star formation in molecular clouds.  
In most cases, the size, mass, and column density of the BGPS sources are similar to large fragments or clumps 
embedded within larger molecular clouds and are often coindident with infrared dark clouds \citep{Peretto:2016}.
Such objects may form small stellar clusters so one might expect that 
these are further fragmented into smaller, less massive but higher density objects that act as seeds for newborn stars. 
The resolution and sensitivity of the LMT data enable an examination of the fragmentation of the dense, massive clumps and 
filaments in molecular clouds distributed over a large range of Galactic radii. 

\begin{figure}
\begin{center}
\epsfxsize=9cm\epsfbox{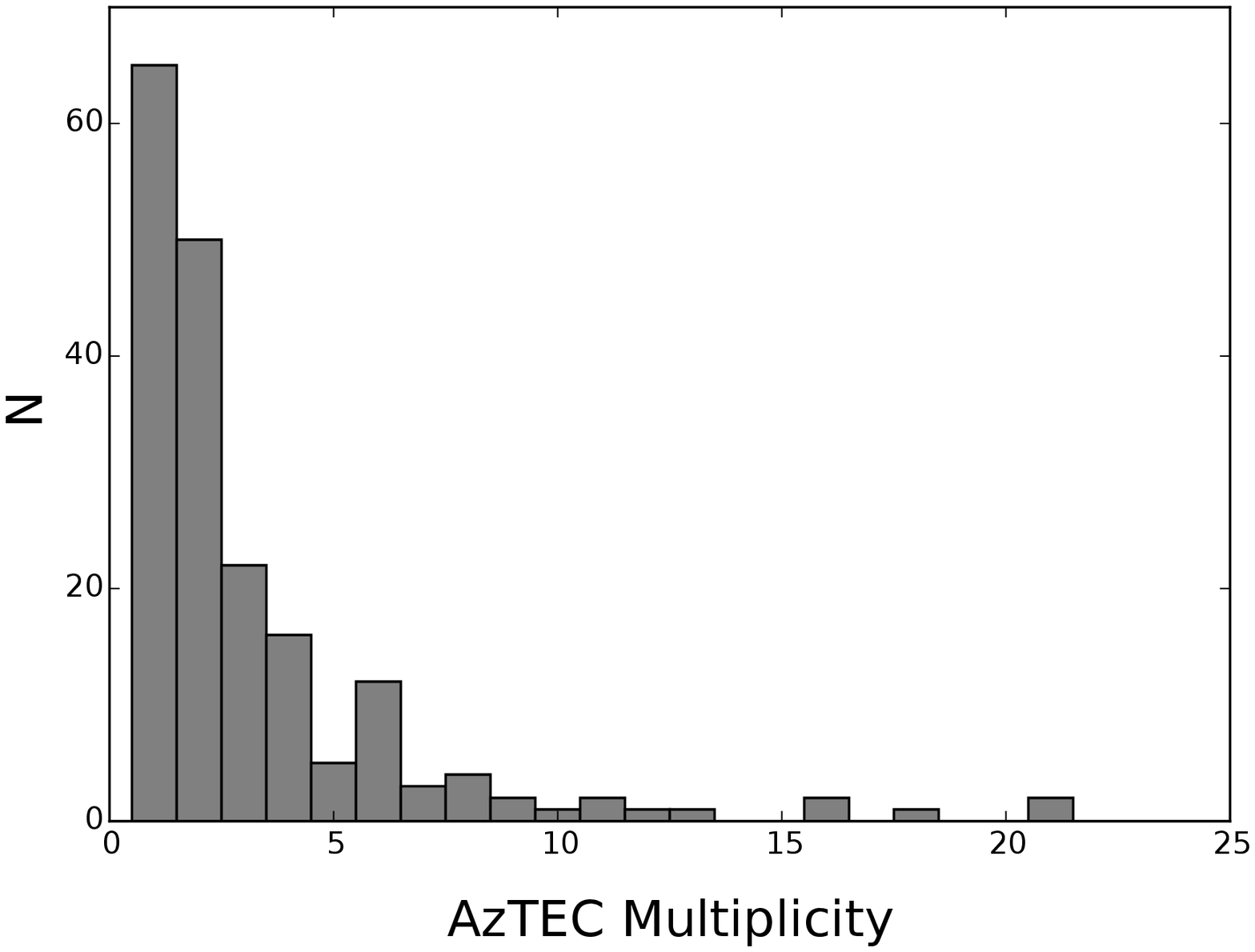}
\caption{Distribution of AzTEC multiplicity in BGPS objects. 
}
\label{fig6}
\end{center}
\end{figure}

We define the AzTEC multiplicity as 
the number of AzTEC sources with $P_J$ values less than 0.2 within a BGPS source boundary.
Figure~\ref{fig6} shows the distribution 
of this multiplicity.  While a single, embedded AzTEC source is the most 
frequent configuration, most BGPS objects contain two or more compact sources.   
In some cases, a BGPS source is highly fragmented 
with greater than 10 AzTEC objects within its domain.  
Possible processes responsible for this clump fragmentation include 
ambipolar diffusion \citep{Lizano:1989}, turbulent fragmentation \citep{Padoan:2002}, and gravitational 
Jeans' instability \citep{Jappsen:2005}.  

While our data can not address the role of magnetic fields and turbulence in 
the fragmentation of clumps, we can evaluate the expected outcome of Jeans' fragmentation in which the mass of 
the resultant sub-fragments are comparable to the Jeans' mass of the larger volume.  
The 
Jeans' mass of a BGPS object is 
\begin{equation}
M_{J,BGPS}=\frac{\pi^{5/2}}{6} G^{-3/2} \rho^{-1/2} c_s^3
\end{equation}
where $c_s=(kT/\mu m_H)^{1/2}$ is the sound speed for isothermal gas, $\rho$ is 
the mean density, and spherical geometry is assumed.
Mean density and temperature values are taken from the compilation by \citet{Svoboda:2016}. 
The mean mass, $<m_{AzTEC}>$, of all embedded AzTEC sources within a BGPS object is calculated.
We also compute the Mach number of non-thermal motions from the measured NH$_3$ velocity dispersion, 
${\sigma}_{v,NT}/c_s=(\sigma_{v,NH3}^2-c(NH_3)^2)^{1/2} / c_s) $, where $c(NH_3)=(kT/17 m_H)^{1/2}$ 
is the thermal speed of ammonia molecules.  Here, we assume the NH$_3$ velocity dispersion from the peak dust 
position is representative of the velocity dispersion over the full clump. 
Values of the Mach number for this subset of BGPS objects range 
from 1.6-7.2.  We caution that the Jeans' criterion for fragmentation, which is based on 
thermal pressure support against self-gravity, may not apply 
to the regime where the gas motions are strongly supersonic.   

\begin{figure}
\begin{center}
\epsfxsize=9cm\epsfbox{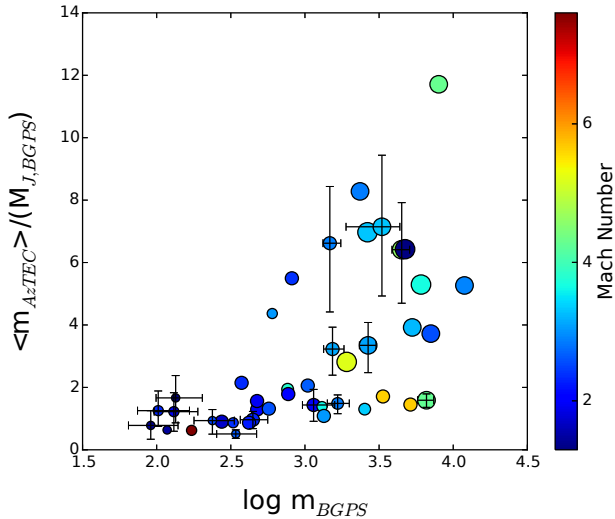}
\caption{The ratio of the mean mass of AzTEC sources to the BGPS Jeans' mass as a function of 
BGPS mass derived from dust emission. The color of the points denote the Mach number for an 
isothermal equation of state.  
The color table ranges between the minimum Mach number of 1.6 to the maximum of 
7.2.  
The size of the points are proportional to the distance 
over the range 3.3~kpc to 13~kpc.  For BGPS sources with distances less than 6 kpc, the 
AzTEC source masses are comparable to the Jeans' mass of the overlying clump.  More distant 
sources exhibit larger values of $<$m$_{AzTEC}>$/M$_{J,BGPS}$.
}
\label{fig7}
\end{center}
\end{figure}

Figure~\ref{fig7} shows 
the variation of the ratio $<m_{AzTEC}>/M_{J,BGPS}$ as a function of the BGPS mass, $m_{BGPS}$, derived 
from the dust emission \citep{Svoboda:2016}.  The colors and sizes of the points represent the Mach number and 
distance respectively.    The results of this analysis are mixed. 
Most of the Galactic near-side (D $<$ 6~kpc), moderate Mach number BGPS objects 
harbor AzTEC fragments with masses comparable to the BGPS Jeans' mass as expected for Jeans' fragmentation. 
Yet, there are also several strongly supersonic BGPS objects with 
$<m_{AzTEC}> \approx M_{J,BGPS}$.   Some of these objects are associated with 
infrared dark clouds \citep{Peretto:2009}
while others are coincident with HII regions that may increase the NH$_3$ velocity dispersion at 
the peak dust position.  
All of the Galactic far-side (D $>$ 9~kpc) BGPS objects in this subsample 
have $<m_{AzTEC}>/M_{J,BGPS}$ greater 
than 3.   This supercritical Jeans' mass state likely arises from the inability of our observations to resolve 
the BGPS Jeans' length in these more distant objects.   In this case, the beam is integrating over 
multiple Jeans' lengths leading to masses in excess of the clump Jeans' mass.  For a gas temperature of 16~K
and a volume density of 1666 \cc\ corresponding to the weighted average of starless and protostellar clumps 
\citep{Svoboda:2016}, a typical BGPS Jeans' length is 0.23~pc.  This scale is not resolved by the 8.5\arcsec\
LMT/AzTEC 
beam at distances greater than 5.5~kpc. 

\subsection{The fraction of dense gas in BGPS clumps}
The correlation of star formation rate with the amount of 
dense gas has been established on the scales of the central regions of galaxies \citep{Gao:2004}, massive 
clumps of molecular clouds \citep{Wu:2005}, and local star forming regions \citep{Heiderman:2010, Lada:2012, 
Gutermuth:2011}.  
In these cases, dense gas 
refers to regions with high visual extinction or molecular line emission that requires
high excitation conditions. Such lines include the rotational transitions of HCN, N$_2$H$^+$, HCO$^+$, and CS with 
critical densities of $\sim$10$^{4}$ \cc\ or higher. 

\begin{figure}
\begin{center}
\epsfxsize=9cm\epsfbox{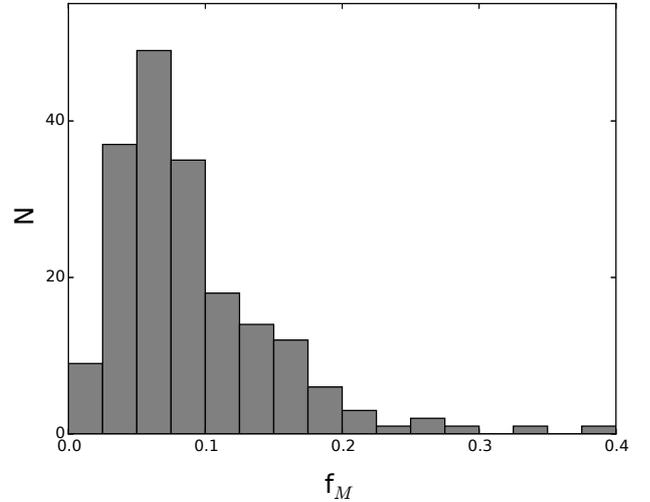}
\caption{Distribution 
of the fraction of $\lambda$1.1~mm flux contributed by embedded AzTEC sources to the total flux measured by the BGPS.
Assuming the same dust temperature between the BGPS and AzTEC sources, 
this ratio corresponds to the fractional mass of the clump distributed into 
small, dense fragments.  
}
\label{fig8}
\end{center}
\end{figure}

To better understand this correlation between star formation rate and the mass of available dense gas, 
a key measure to evaluate is the fraction of a clump's mass 
that is distributed into subfragments at higher density as 
these provide the local reservoir of gas that fuels the current or impending formation of stars.
Assuming the same dust temperature for the BGPS object and its linked AzTEC 
sources, this fractional mass, $f_{M}$, is directly calculated from the ratio of total flux from all AzTEC sources to the 
flux measured by the BGPS, 
\begin{equation}
f_{M}=\frac{\Sigma M_{i,AzTEC}} {M_{BGPS}}=\frac{\Sigma S_{i,AzTEC}}{S_{BGPS}}
\end{equation}
where the sum is over all AzTEC sources within the BGPS source. 
Figure~\ref{fig8} shows the distribution of $f_{M}$ for all BGPS sources linked to 1 or more AzTEC sources.  
The median value is 0.075 with the standard deviation of 0.059 and a mean, 1$\sigma$ measurement uncertainty of 0.026.   
This fraction does not include isolated, 
low mass regions that are external to the compiled AzTEC sources but whose dust emission level 
falls below our 
sensitivity limit.  The contribution  of low mass fragments or cores 
to the total clump mass is included in the BGPS integrated flux measure.  Therefore, this calculation of 
the dense gas fraction could be affected by incompleteness of AzTEC to detect low mass objects.
However, we find no decreasing trend 
in $f_M$ values with increasing distance as one would expect as low-mass sources fall below our mass sensitivity limit.
The absence of such a trend suggests that most low mass, compact sources are primarily contained 
within the solid angle of the extracted AzTEC sources so that their contributions are included and our calculation of the 
fractional mass of dense gas is not severely underestimated.

Given the mass distribution of AzTEC sources, these are not representative of protostellar cores that may collapse into 
single  stars.  Rather, the AzTEC sources are likely further structured into even denser 
gas subfragments as shown by sub-arcsecond interferometry of star forming regions over fields comparable 
to the AzTEC beam solid angle \citep{Beuther:2007, Brogan:2016, Cyganowski:2017}.  Thus, the fraction of 
the clump mass distributed into denser, protostellar cores is likely much smaller than 8\%.

The BGPS clumps reside within the larger structures of molecular clouds that are 
typically probed by CO and its isotopologues
\citep{Heyer:2015}. 
From a sample of over 300 clouds, \citet{Battisti:2014} determined the fractional mass of a cloud distributed into 
BGPS sources 
is $f_B$=11\% with a population dispersion of 6\%.  For nearby star forming regions, \citet{Lada:2012} derived a 
similar fractional mass for regions with A$_v >$ 8 magnitudes.
Taking this value for the fractional mass of BGPS sources in molecular clouds within our survey field, then AzTEC 
sources comprise 0.008$\pm$0.008 of a cloud mass.  Protostellar cores within the AzTEC source constitute an even 
smaller fraction.  These results emphasize that within time intervals smaller than the dynamical time 
of AzTEC sources, 
only a small fraction of a molecular cloud is configured for the production of new stars. 
\section{High mass star formation thresholds}
The formation of a high mass (m$_* >$ 8~\msun) star within the ISM is a rare event, which implies that 
restrictive conditions are required.  For the core accretion description of high mass star formation, 
the protostellar core must have sufficient 
initial mass to produce a massive star and a sufficient accretion rate to overcome radiation pressure 
\citep{McKee:2003}. 
In addition, fragmentation of the core must be suppressed to allow for most of the mass to accrete onto a 
singular, developing 
star.  
Radiative heating from recently formed low mass stars can increase the Jeans' mass of the local core and 
suppress any further 
fragmentation.  \citet{Krumholz:2008} derive a mass surface density threshold, $\Sigma_{\rm cr}$, at which the 
light-to-mass ratio of accreting, low mass protostars is equal to light-to-mass ratio 
necessary to 
suppress fragmentation.  
For solar metallicity and a background dust temperature of 10~K, $\Sigma_{\rm cr}$ 
varies between 0.7 and 1 g cm$^{-2}$
for stellar masses between 10 and 100~\msun.  An empirically defined mass threshold of clumps was identified 
by \citet{Kauffmann:2010} for a set of 
infrared dark clouds with radius r, 
$m_{cl,min}$=870 \msun (r/pc)$^{1.33}$ 
that separates IRDCs with and without signatures of massive star formation.  
This minimum mass corresponds to a 
mass surface density threshold of 0.05 g cm$^{-2}$ (r/1~pc)$^{-0.67}$. \citet{Urquhart:2014} identified a similar 
mass surface density threshold based on a sample of ATLASGAL clumps linked to various tracers of high mass star formation. 

\begin{figure*}
\begin{center}
\epsfxsize=13cm\epsfbox{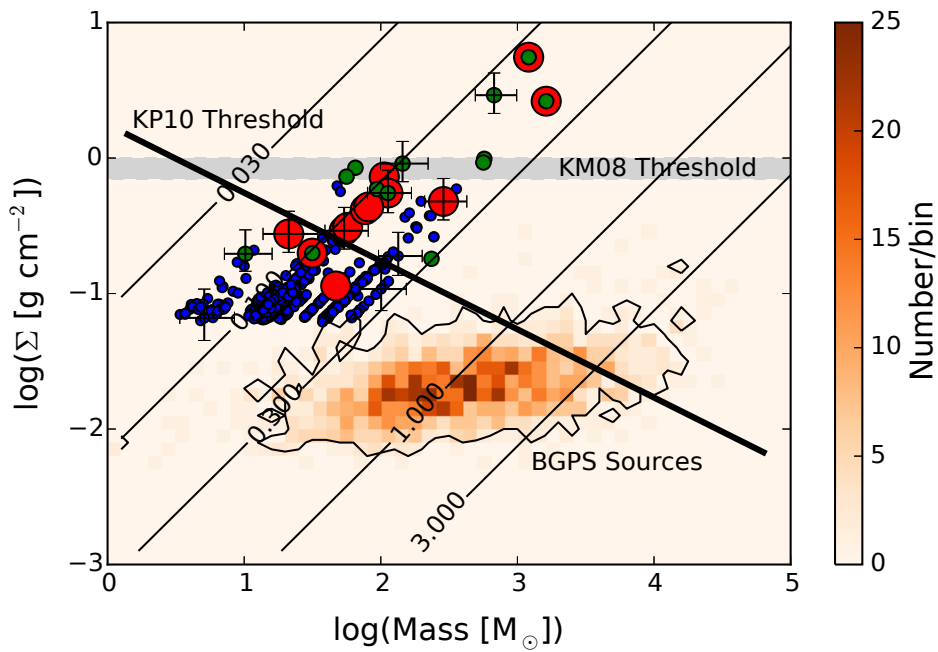}
\caption{Distribution of AzTEC sources in the mass-$\Sigma$ plane linked to UCHII regions (red), methanol masers (green), 
and no signatures to massive star formation (blue).  Since the AzTEC sources are unresolved or marginally resolved, 
the $\Sigma$ values are lower limits to structures witin the AzTEC 8.5\arcsec\ beam. 
The halftone is a 2D histogram of mass and $\Sigma$ values
for 1640 BGPS sources compiled by \citet{Svoboda:2016} with established masses.  The horizontal grey wedge 
marks the mass surface density threshold 
proposed by \citet{Krumholz:2008} and the solid back line represents the empirical threshold of \citet{Kauffmann:2010}.
For clarity, every 10th errorbar is plotted.  
Labeled diagonal lines are loci of constant radius in parsecs, r=(Mass/$\pi\Sigma$)$^{1/2}$.
}
\label{fig9}
\end{center}
\end{figure*}

To further investigate this proposed mass surface density threshold, we examine our sample of AzTEC sources.
Figure~\ref{fig9} shows the location 
of the identified AzTEC objects within the mass-$\Sigma$ plane that is frequently used to 
distinguish the environments of low and high mass star formation \citep{Tan:2014}.  Sources are differentiated 
by 
their association with UCHII regions (red points), Class II methanol masers (green points), 
 or absence of either of these signatures
(blue points).  Several AzTEC sources harbor both a compact HII region and a Class~II methanol maser. 
The large sample of  BGPS sources analyzed by \citet{Svoboda:2016} with derived masses are represented 
as a 2-dimensional histogram.
The diagonal lines show loci of constant radius in parsecs.
The mass surface density thresholds to massive star formation 
proposed by \citet{Krumholz:2008} and \citet{Kauffmann:2010} are shown as the grey wedge and solid black line 
respectively.  We re-emphasize that the derived AzTEC mass surface densities and source sizes are lower and upper limits 
respectively since 
these sources are marginally resolved or unresolved.  The apparent 
correlation between mass and mass surface density of the AzTEC points or equivalently, equal source size, 
 is a result of adopting a common 
solid angle, $\Omega_A$, for each source. 
The parallel bands of points 
correspond to distinct clusters of 
AzTEC sources located at different distances (see Figure~\ref{fig4}).

Most of the AzTEC sources associated with ultra-compact HII regions are located above the KP2010 threshold that affirms 
this empirically derived limit.   
More interestingly, 10 of the 13 AzTEC sources linked to Class II methanol masers are positioned at or above 
the more constraining KM08 threshold within the measurement errors.  This strong bias offers provisional evidence for high mass 
star forming regions satisfying the restrictive conditions that suppress core fragmentation.
To more fully evaluate these conditions, higher angular 
resolution measurements are required that can discern whether one or more objects are present within these AzTEC sources. 
A single, massive fragment identified within an AzTEC source with these  mass surface density values or higher would provide 
powerful evidence for the suppression of fragmentation that leads to the formation of a massive star.

\section{Conclusions}
Sensitive, imaging observations with 8.5\arcsec\ angular resolution of the 1.1mm dust continuum 
emission along the Galactic plane have been used to 
examine the properties of high density fragments and subfragments within molecular clouds.  The identified 
objects are embedded within larger clumps of molecular clouds as measured by the Bolocam Galactic Plane Survey.  
The AzTEC sources have larger column and mean volume densities and 
smaller masses than the overlying clumps but may also be further fragmented when observed with higher angular resolution.  
The AzTEC sources comprise on average, 8\% of the mass of the BGPS clump in which these reside and 0.8\% of the mass of the 
parent molecular cloud.   
The embedded AzTEC sources have masses that are a factor of 1-10 times larger than the Jeans' mass
of the BGPS object. 
The mass surface  densities of AzTEC sources linked to 
ultra-compact HII regions 
are typically greater than the mass surface density implied by the empirical mass-size threshold \citep{Kauffmann:2010}.  For AzTEC sources associated with 
Class II methanol masers 
that mark  an 
earlier phase of massive star formation, the mass surface densities exceed 
 0.7  g cm$^{-2}$ that approaches the critical threshold  proposed by \citet{Krumholz:2008}.

\section*{Acknowledgments}
We appreciate the useful comments provided by referee that improved the clarity of this manuscript. 
This research made use of Astropy, a community-developed core Python package for Astronomy (Astropy Collaboration, 2013).
We also acknowledge the long-term financial support from the Mexican Science and Technology Funding Agency, 
Consejo Nacional de Ciencia y Tecnolog\'ia (CONACYT), during the construction and early science phase of the 
Large Millimetre Telescope Alfonso Serrano, as well as support from the US National Science Foundation via the 
University Radio Observatory program, the Instituto Nacional de Astrof\'isica, \'Optica y Electronica (INAOE), 
and the University of Massachusetts (UMASS).

\bibliographystyle{mnras}
\bibliography{cite_g24} 

\end{document}